\documentclass[nofootinbib]{revtex4}
\usepackage{CJK,upgreek,fancyhdr}
\usepackage{graphicx}
\usepackage{latexsym}
\usepackage{amsmath}

\def\be{\begin{equation}}
\def\ee{\end{equation}}
\def\bea{\begin{eqnarray}}
\def\eea{\end{eqnarray}}

\begin{document}
\begin{CJK*}{UTF8}{gbsn}

\hfill USTC-ICTS-19-25

\title{Eddington-inspired Born-Infeld Gravity with Varying Cosmological Constant}

\author{Haomin Rao}
\email{rhm137@mail.ustc.edu.cn}
\author{Dehao Zhao}
\email{dhzhao@mail.ustc.edu.cn}
\affiliation{Interdisciplinary Center for Theoretical Study, University of Science and Technology of China}


\begin{abstract}
 In this paper we modify the EiBI model to realize a varying cosmological constant which is determined by matter distribution.
 We find that the Newton's constant is also variable and its change is related to the change of cosmological constant.
 And then we study its cosmological behavior.
 We find that the early universe will have different behaviors if we take different forms of pending functions.
 And we can avoid singularity in early universe just like the original EiBI model.
\end{abstract}

\maketitle


\section{Introduction}

Among various dark energy candidates, the cosmological constant $\Lambda$ is phenomenologically the simplest one and gives well account of observational data, even though it brought lots of puzzles challenging the fundamental physics. In general relativity (GR), $\Lambda$ is forced to be a true constant by the covariance of the theory. However, people have continues interests to know whether the cosmological constant can be promoted to be a dynamical variable. One way to promote the cosmological constant is replacing $\Lambda$ by a dynamical scalar field with kinetic and potential terms. In this way one introduces a new propagating degree of freedom to the universe, as people have done in quintessence or some other dynamical dark energy models. Another way is treating $\Lambda$ as a variable without kinetic term in the action, like the Lagrange multiplier. However, naively promote $\Lambda$ as a Lagrange multiplier in the action of GR$+\Lambda$, the variation with respect to $\Lambda$ yields vanishing spacetime volume: $\sqrt{-g}=0$, where $g$ is the determinant of the metric tensor $g_{\mu\nu}$. This is physically nonsense.

Recently, a model with varying cosmological constant using the second approach mentioned above was proposed in Ref. \cite{VaryingLambda} by adding a ``quasi-topological" term to the action of GR$+\Lambda$:
\be
S_{quasi}=-\int d^4x \sqrt{-g} \frac{3}{2\Lambda}\mathcal{C}~,
\ee
where $\sqrt{-g}\mathcal{C}$ is a topological invariant, such as the Gauss-Bonnet term or the Pontryagin invariant. Because this term is inversely proportional to $\Lambda$, contrary to the original $\Lambda$ term,  the variation with respect to $\Lambda$ gives the result that $\Lambda^2$ is proportional to the topological invariant whose spacetime dependence is further determined by the matter distribution and the dynamical torsion in the universe. This model has some consequences on cosmology, this was studied in Ref. \cite{Alexander:2019wne}.

We note that, to realize a varying cosmological `constant', $\Lambda$ was inserted into the denominator of the ``quasi-topological" term in Ref. \cite{VaryingLambda} . We also note that another case where $\Lambda$ appears in the denominator is the action of the Eddington gravity \cite{eddington}:
\be
S_{Edd}=\frac{1}{\Lambda}\int d^4x \sqrt{|R|}~,
\ee
where $|R|$ is the determinant of the Ricci tensor $R_{\mu\nu}$, which was purely determined by the affine connection $\Gamma^{\rho}_{\,\mu\nu}$. The Eddington gravity is equivalent to GR in the absence of matter. But it has difficulty to account for matter-gravity couplings, since there is no metric at the beginning. Historically, Eddington's gravity has been extended several times \cite{BI1988,BI2004,BI2005}. And several years ago,
Eddington¡¯s gravity was expanded again in Ref. \cite{EIBI}, where the action has the Born-Infeld like structure:
\begin{equation}\label{EIBI}
  S_{B I}[g, \Gamma, \Psi]=\frac{1}{\kappa} \int d^{4} x\left[\sqrt{-\left|g_{\mu \nu}+\kappa R_{\mu \nu}(\Gamma)\right|}-\lambda \sqrt{-g}\right]+S_{M}[g, \Psi]~,
\end{equation}
where $\Psi$ is matter field. This so-called Eddington-inspired Born-Infeld (EiBI) gravity is not purely affine, the metric tensor $g_{\mu\nu}$ was introduced at the starting point. In this theory, the matter action and couplings were added in the same way as in GR. The EiBI approaches to the model of GR$+\Lambda$ for small values of $\kappa R$, where the cosmological constant is
\be\label{EIBILambda}
\Lambda=\frac{\lambda-1}{\kappa}~,
\ee
but it deviates GR significantly for large curvature.

In this paper we will apply the idea of Ref. \cite{VaryingLambda} to the EiBI model to realize a varying cosmological constant in EiBI. Since in the original EiBI model \cite{EIBI}, the cosmological constant $\Lambda$ was determined by two constants $\kappa$ and $\lambda$ as shown in Eq. (\ref{EIBILambda}), we will promote both of them to be dynamical variables in our new model. Generally we will replace them by the functions of $\kappa$ as in the following action:
\begin{equation}
S[g, \Gamma, \kappa, \Psi]=\int d^{4} x \frac{1}{f_{1}(\kappa)} \left[\sqrt{-\left|g_{\mu \nu}+f_{3}(\kappa) R_{\mu \nu}(\Gamma)\right|}-f_{2}(\kappa) \sqrt{-g}\right]+S_{M}[g, \Psi]~.
\end{equation}
Please note that here we allow that the functions $f_{1}(\kappa)$ and $f_3(\kappa)$ may be different. In the original EiBI, both of them are replaced by the same constant $\kappa$. Since all the $f_i(\kappa)$ functions are dependent on $\kappa$, we can always make $f_{3}(\kappa)=\kappa$ by redefinition. We will do so hereafter and start with the simpler action:
\begin{equation}\label{action}
S[g, \Gamma, \kappa, \Psi]=\int d^{4} x \frac{1}{f_{1}(\kappa)} \left[\sqrt{-\left|g_{\mu \nu}+\kappa R_{\mu \nu}(\Gamma)\right|}-f_{2}(\kappa) \sqrt{-g}\right]+S_{M}[g, \Psi]~.
\end{equation}

\section{Theory and general analysis}

We will study the model (\ref{action}) in terms of the first order formalism in which the metric $g_{\mu\nu}$ and the connection
$\Gamma^{\rho}_{\,\mu\nu}$ are considered as independent variables when using the variational principle. For the sake of simplicity, we do not consider the torsion and assume that the matter does not couple to the connection $\Gamma^{\rho}_{\,\mu\nu}$ directly.
The equations of motion can be obtained by varying the action with respect to  $g_{\mu\nu}$, $\Gamma^{\rho}_{\,\mu\nu}$  and $\kappa$ respectively:
\bea
& &\frac{\sqrt{-\left|Q\right|}}{\sqrt{-g}}(Q^{-1})^{\mu \nu}-f_{2} g^{\mu \nu}+f_{1} T^{\mu \nu}=0~,\label{eom1}\\
& &\nabla_{\rho}\left[\frac{\kappa}{f_{1}} \sqrt{-\left|Q\right|}\left(Q^{-1}\right)^{\mu \nu}\right]=0~,\label{eom2}\\
& & \frac{\sqrt{-\left|Q\right|}}{\sqrt{-g}}\left(\frac{f_{1}}{2}(Q^{-1})^{\mu \nu}R_{\mu\nu}-f'_{1}\right)+f'_{1}f_{2}-f'_{2}f_{1}=0~,\label{eom3}
\eea
where $Q_{\mu\nu}= g_{\mu\nu}+\kappa R_{\mu\nu}$, $\left|Q\right|$ is the determinant of $Q_{\mu\nu}$, $\left(Q^{-1}\right)^{\mu \nu}$ is the inverse matrix of $Q_{\mu\nu}$, $T^{\mu \nu}$
is the standard energy-momentum tensor, and the prime represents the derivative with respect to $\kappa$.
Eq. (\ref{eom3}) is a constraint equation, it means that the $\kappa$ is determined by $g_{\mu\nu}$ and $R_{\mu\nu}$. Other equations of motion tell us that $g_{\mu\nu}$ and $R_{\mu\nu}$ are determined by the energy-momentum tensor and $\kappa$.  Roughly speaking, $\kappa$ is determined by matter distribution.

From Eq. (\ref{eom2}), we know that there is an auxiliary metric $q_{\mu\nu}\equiv (\kappa/f_1)Q_{\mu\nu}$ which is compatible with the connection
$\Gamma^{\rho}_{\,\mu \nu}$, i.e., $\Gamma^{\rho}_{\,\mu \nu}=\frac{1}{2} q^{\rho \sigma}\left(q_{\mu \sigma, \nu}+q_{\nu \sigma, \mu}-q_{\mu \nu, \sigma}\right)$ and $q^{\mu\nu}$ is the inverse of $q_{\mu\nu}$.
In terms of the auxiliary metric, we can rewrite the above three equations of motion as
\bea
& &\frac{f_{1}}{\kappa}\frac{\sqrt{-q}}{\sqrt{-g}}q^{\mu \nu}-f_{2}g^{\mu \nu}+f_{1} T^{\mu \nu}=0~,\label{Eom1}\\
& &g_{\mu\nu}+\kappa R_{\mu\nu}=\frac{f_{1}}{\kappa}q_{\mu\nu}~,\label{Eom2}\\
& &\frac{\sqrt{-q}}{\sqrt{-g}}\left(\frac{f_{1}}{\kappa}\right)^{2}\left(\frac{\kappa}{2}  q^{\mu \nu} R_{\mu \nu}-f'_{1}\right)+\left(f'_{1} f_{2}-f'_{2} f_{1}\right)=0~,\label{Eom3}
\eea
where $q$ is the determinant of $q_{\mu\nu}$.

Similar to the original EiBI model \cite{EIBI}, our extended gravity model should approach GR with a varying cosmological constant at the the regime of small curvature, because GR has been tested by many experiments at low energy scales. Now we will investigate how this requirement constrains our model.
At the limit of small curvature where $\kappa R_{\mu\nu}\ll g_{\mu\nu}$, one obtains
\be
q^{\mu\nu}\simeq {f_1 \over \kappa}(g^{\mu\nu}-\kappa R^{\mu\nu})~,~~\frac{\sqrt{-q}}{\sqrt{-g}}\simeq \frac{\kappa^2}{f_1^2}(1+\frac{1}{2}\kappa R)
\ee
in terms of Eq. (\ref{Eom2}), where $R^{\mu\nu}=g^{\mu\alpha}g^{\nu\beta}R_{\alpha\beta}$ is the contravariant Ricci tensor and $R$ is its trace, i.e., the curvature scalar. With these, the Eq. (\ref{Eom1}) at the first order of $\kappa R$ becomes
\be\label{einstein}
G^{\mu\nu}\simeq \frac{f_1}{\kappa}T^{\mu\nu}-\frac{f_2-1}{\kappa} g^{\mu\nu}~.
\ee
We can see that the gravitational field equation (\ref{einstein}) at low curvature scale is very similar to the Einstein equation of GR, except that both Newton's `constant' $G=(1/8\pi)(f_1/\kappa)$ and the cosmological `constant' $\Lambda=(f_{2}-1)/\kappa$ are generally spacetime dependent variables. At the same order the constraint equation (\ref{Eom3}) is approximately
\be\label{constraint}
d(\frac{f_2-1}{\kappa})\simeq (\frac{f_1-1}{\kappa}-\frac{R}{2})d\ln (\frac{f_1}{\kappa})~, ~{\rm or}~d\Lambda\simeq (\Lambda-\frac{R}{2} )d\ln G~.
\ee
This equation showed how Newton's and the cosmological `constants' change with the spacetime. Because Eq. (\ref{constraint}) relies on the curvature scalar, these two `constants' are finally determined by the matter distribution.
In fact, take the trace of Eq. (\ref{einstein}) and substitute it into Eq. (\ref{constraint}), one will obtain the following relation,
\be\label{constraint2}
d\Lambda=(4\pi G T-\Lambda)d\ln G~,
\ee
where $T$ is the trace of the energy-momentum tensor.
What should be noted is the special case where $f_1=\kappa$. In this case Eq. (\ref{constraint}) showes that $d\Lambda=0$, so both $G$ and $\Lambda$ are true constants even though $\kappa$ itself is a spacetime function, and the model returned to GR$+\Lambda$ exactly.

Besides this special case, the Eqs. (\ref{constraint}) or (\ref{constraint2}) impose the restriction that in this model these two `constants' should vary together. A varying Newton's `constant' has been studied a lot in the Ref. \cite{Uzan:2010pm}, especially in the framework of scalar-tensor theories of gravity. Currently there are some upper limits on the change rate of Newton's `constant' imposed by experiments or observations, for instances, the constraints from pulsars \cite{Williams:2004qba}, lunar laser ranging \cite{Kaspi:1994hp,Zhu:2018etc}, Big Bang nucleosynthesis \cite{Copi:2003xd,Cyburt:2004yc} are roughly at the same order: $|d\ln G/dt|_{t_0}\lesssim 10^{-12}~{\rm year}^{-1}$.  These results also put a constraint on the change rate of the cosmological `constant' through the model discussed here.
In the late universe (low curvature and $\Lambda$ dominant) of the $\Lambda$CDM cosmology, the equation (\ref{constraint2}) implies
\be\label{Glimitation}
\left|\frac{d\ln \Lambda}{dt}\right|_{t_0}\sim \left|\frac{d\ln G}{dt}\right|_{t_0}\lesssim 10^{-12}~{\rm year}^{-1}~.
\ee
Locally, Newton's `constant' is spatial dependent, its variation depends on local matter distribution. Given this, we may know how the cosmological `constant' vary with spatial position in terms of Eqs. (\ref{constraint}) or (\ref{constraint2}).

As in the original EiBI model, our extended model deviates GR significantly at the high curvature scales, $\kappa R_{\mu\nu}\gg g_{\mu\nu}$.
At these scales, $q_{\mu\nu}\simeq (\kappa^2/f_1) R_{\mu\nu}$, the action of gravity approaches Eddington's except the difference that $\kappa$ and $f_1$ are functions instead of constants. The Eq. (\ref{Eom1}) dictates how the auxiliary metric $q_{\mu\nu}$ which is compatible with the connection relates to the physical metric $g_{\mu\nu}$ to which the matter couples minimally, this will transform to the relation between the curvature and matter,
\be
\kappa \sqrt{-|R|}(R^{-1})^{\mu\nu}\simeq \sqrt{-g}(f_2g^{\mu\nu}-f_1 T^{\mu\nu})~,
\ee
where $(R^{-1})^{\mu\nu}$ is the inverse of $R_{\mu\nu}$. The constraint equation (\ref{Eom3}) in this limit becomes,
\be
\kappa \sqrt{-|R|}(2f_1-\kappa f_1')\simeq \sqrt{-g}(f_2'f_1-f_1'f_2)~.
\ee
Combined these two equations, our model yields the following relation at the high curvature scales:
\be
(R^{-1})^{\mu\nu}\simeq \frac{2f_1-\kappa f_1'}{f_2'f_1-f_1'f_2}(f_2g^{\mu\nu}-f_1 T^{\mu\nu})~.
\ee
This is quite different from Einstein's equation and different behaviors of the system governed by this gravity model is strongly expected at high curvature regime. Below we will apply this model to cosmology. One may expect that, in comparison with the standard cosmology based on GR,  significant difference appears at the early universe and diminishes to a negligible level at late times. We will confirm these results by solving Eqs. (\ref{Eom1})(\ref{Eom2})(\ref{Eom3}) numerically in the universe.

By the way, for the original EiBI model, there is an equivalent bimetric-like action \cite{NewInsight}.
Similarly, there is an equivalent bimetric-like action in our model,
\begin{equation}\label{newaction}
S[g, q, \kappa, \Psi]=\int d^{4} x \{\sqrt{-q}\left[ \frac{q^{\mu\nu}R_{\mu\nu}(q)}{2}+\frac{1}{2 \kappa} q^{\mu \nu} g_{\mu \nu}-\frac{f_{1}}{\kappa^{2}}\right]-\frac{f_{2}}{f_{1}} \sqrt{-g}\}+S_{M}[g, \Psi]~.
\end{equation}
One can easily prove that the equations of motion obtained from variations of above bimetric-like action with respect to $g_{\mu \nu}, q_{\mu \nu}$ and $\kappa$
are equivalent to Eqs. (\ref{Eom1})(\ref{Eom2})(\ref{Eom3}).
Therefore the action (\ref{newaction}) is equivalent to the original action (\ref{action}).

\section{Application to Cosmology}

\subsection{Homogeneous and isotropic background}
   Imitating the discussion in Ref.~\cite{EIBI}, we apply this theory to cosmology. We assume that the background is Friedmann-Robertson-Walker (FRW) spacetime, which is homogeneous and isotropic. And we only consider the case that the space is flat. So the metric $g_{\mu\nu}$ and the auxiliary metric $q_{\mu\nu}$ have the following form
\bea
& &ds^{2}=g_{\mu\nu}dx^{\mu}dx^{\nu}=-dt^{2}+a^{2}(t)\delta_{ij}dx^{i}dx^{j}~,\label{gFRW}\\
& &q_{\mu\nu}dx^{\mu}dx^{\nu}=-U(t)dt^{2}+V(t)a^{2}(t)\delta_{ij}dx^{i}dx^{j}~.\label{qFRW}~.
\eea
The ansatz for $q_{\mu\nu}$ in terms of  two functions $U(t)$ and $V(t)$ was first adopted in Ref. \cite{EIBI}.
We also assume that the matter can be considered as perfect fluid and its energy-momentum tensor has the form: $T^{\mu \nu}=(p+\rho) u^{\mu} u^{\nu}+p g^{\mu \nu}$.

From Eq. (\ref{Eom1}), we can get
\begin{equation}\label{UV}
  U=\frac{\kappa}{f_{1}}\frac{D}{f_{2}+f_{1}\rho}\quad \text { and } \quad V=\frac{\kappa}{f_{1}}\frac{D}{f_{2}-f_{1}p}~,
\end{equation}
where $D=\sqrt{(f_{2}+f_{1}\rho)(f_{2}-f_{1}p)^3}~$.
Then Eq. (\ref{Eom2}) becomes
\begin{equation}\label{Feq}
  (H+\frac{\dot{V}}{2V})^{2}=\frac{1}{6\kappa}
  \left(1+2\frac{f_{1}}{\kappa}U-3\frac{U}{V}\right)~,
\end{equation}
where $\displaystyle{H=\dot{a}/a}$ and dot represents derivative with respect to time.
These equations can be returned to the equations in Ref. \cite{EIBI} when we set $f_{1}=\kappa=\text{constant}$ and $f_{2}=\lambda=$ constant.
The constraint equation (\ref{Eom3}) has the following form in the FRW universe:
\begin{equation}\label{yueshu1}
    2f_{1}f_{2}+\kappa(f_{2}'f_{1}-f_{1}'f_{2})+
   (\kappa f_{1}'-2f_{1})D=\frac{1}{2}f_{1}^{2}T~,
 \end{equation}
where $T=g_{\mu\nu}T^{\mu\nu}=-\rho+3p$ is the trace of energy-momentum tensor.
In addition, from the diffeomorphism invariance of the matter action $S_{M}[g, \Psi]$, we can get the continuity equation
\begin{equation}\label{continuous}
  \dot{\rho}+3H(\rho+p)=0~.
\end{equation}

 \subsection{The behavior of very early universe}
 From discussions in previous section we know that the model is consistent with Einstein's gravity at small curvature scales (low energy density).
 So the cosmological evolution equations are close to those of standard cosmology at late time.
In the very early universe, however, the energy density and the curvature are large enough. The model seriously deviates from Einstein's gravity so that
 the evolution of universe will seriously deviate from the standard big-bang cosmology.
 So we will focus on the very early universe under this model.

For this purpose, we will assume that the very early universe is dominated by radiation, $p=\rho/3$ and $\rho\propto a^{-4}$ due to the continuity equation (\ref{continuous}).
For the convenience of analysis, we define an auxiliary density as $\bar{\rho}= f_{1}\rho/f_{2}$.
Then, from Eq. (\ref{Feq}) together with Eqs. (\ref{yueshu1}) and (\ref{continuous}), we can get
 \begin{equation}\label{3HH}
   H^{2}=\frac{\bar{\rho}-1+\frac{{f_{2}}}{3\sqrt{3}}\sqrt{(1+\bar{\rho})(3-\bar{\rho})^{3}}}
   {3\kappa\left[3+\bar{\rho}^{2}(1-\mathcal{F})\right]^2}(1+\bar{\rho})(3-\bar{\rho})^{2}~,
 \end{equation}
 where
 \be\nonumber \mathcal{F}=\frac{\alpha(3-\bar{\rho})^{\frac{1}{2}}+\gamma(1+\bar{\rho})(3-\bar{\rho})^{\frac{3}{2}}}
{\beta(3-\bar{\rho})^{\frac{1}{2}}+\psi(1+\bar{\rho})^{\frac{1}{2}}+\phi(1+\bar{\rho})(3-\bar{\rho})^{\frac{3}{2}}}~,\ee
and
\bea
& &\nonumber\alpha=4\kappa f_{2}(\kappa f'_{1}-2f_{1})\left[f_{1}f'_{2}(3+\bar{\rho})+f_{2}f'_{1}\bar{\rho}(1-\bar{\rho})\right],
\ \gamma=4f_{2}^{2}(\kappa f'_{1}-2f_{1})(f_{1}-\kappa f'_{1})~,\\
& &\nonumber\beta=2\kappa f_{2}(\kappa f'_{1}-2f_{1})\left[f_{1}f'_{2}(3+2\bar{\rho})-f_{2}f'_{1}\bar{\rho}^{2}\right]~,
\ \phi=\kappa f_{2}^{2}\left[(f_{1}-\kappa f'_{1})f'_{1}+\kappa f_{1}f''_{1}\right]~,\\
& &\nonumber\psi=3\sqrt{3}\kappa\left[2f_{1}^{2}f'_{2}+(f_{1}-\kappa f'_{1})(f'_{2}f_{1}-f'_{1}f_{2})+\kappa f_{1}(f''_{2}f_{1}-f''_{1}f_{2})\right]~.
\eea

Then, solve Eq. (\ref{3HH}) together with Eqs. (\ref{yueshu1}) and (\ref{continuous}),
one can determine how the scale factor evolves over time.
This depends on the function forms of $f_{1}(\kappa)$ and $f_{2}(\kappa)$.

\begin{figure}[h]
  \centering
  \includegraphics[width=0.5\textwidth]{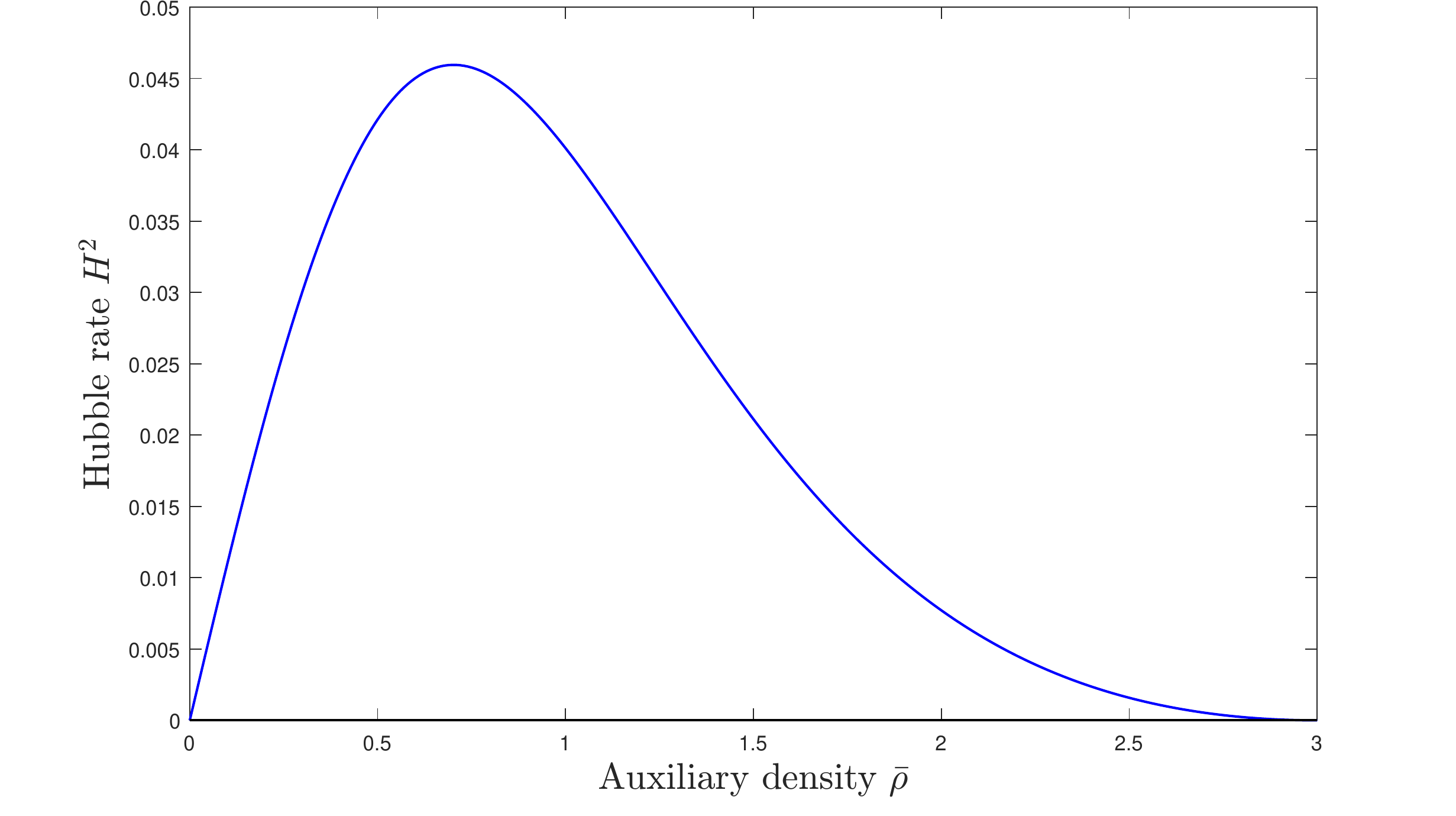}\includegraphics[width=0.5\textwidth]{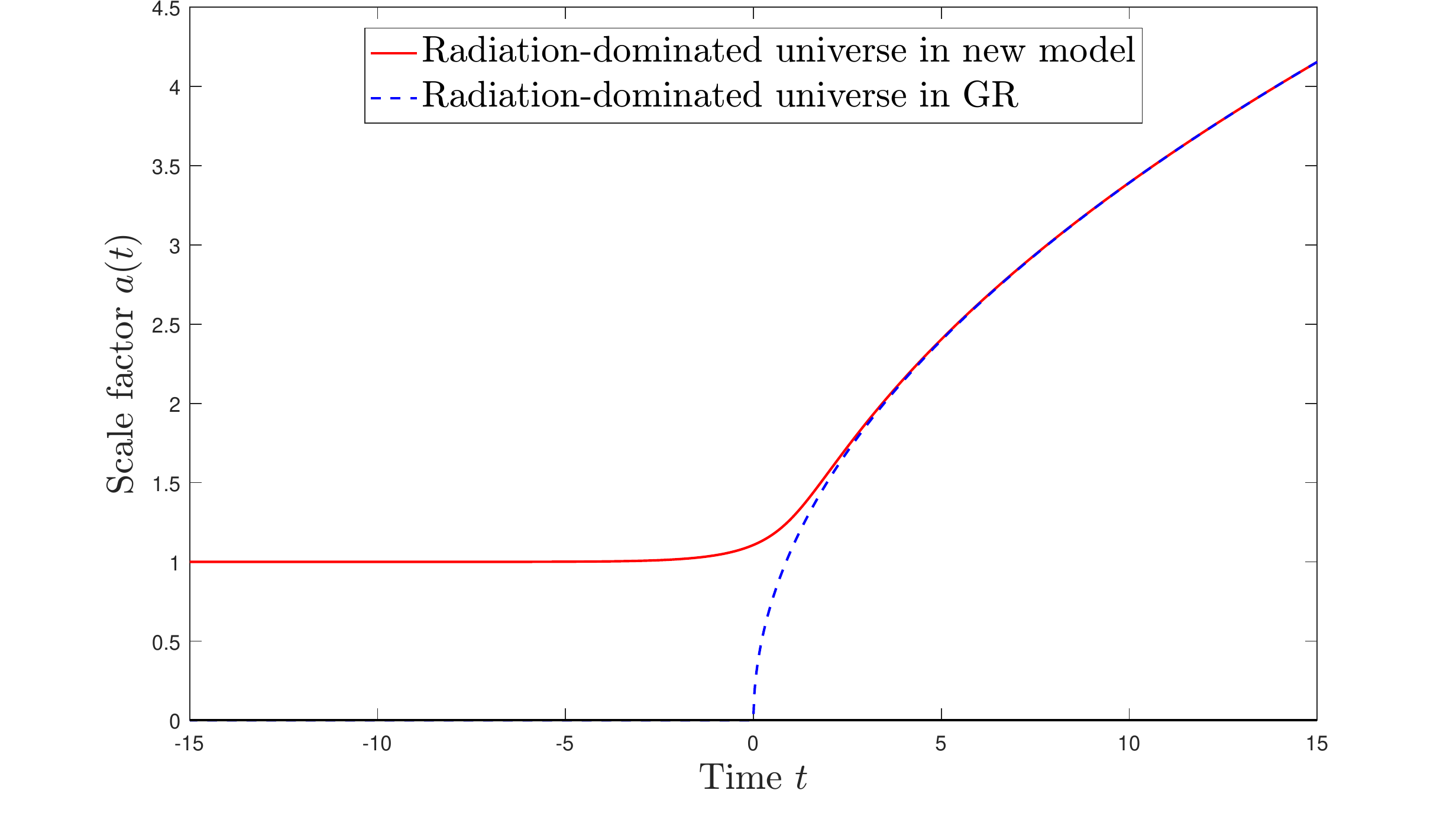}\\
  \caption{Left: Hubble rate $H^{2}$ against auxiliary density in case (1)\\
  Right: Scale factor (normalized by minimum scale factor) against time in case (1). The red solid line represents the evolution in our model
  and the blue dotted line represents the evolution of GR.}
  \label{figure1}
\end{figure}

\begin{figure}[h]
  \centering
  \includegraphics[width=0.5\textwidth]{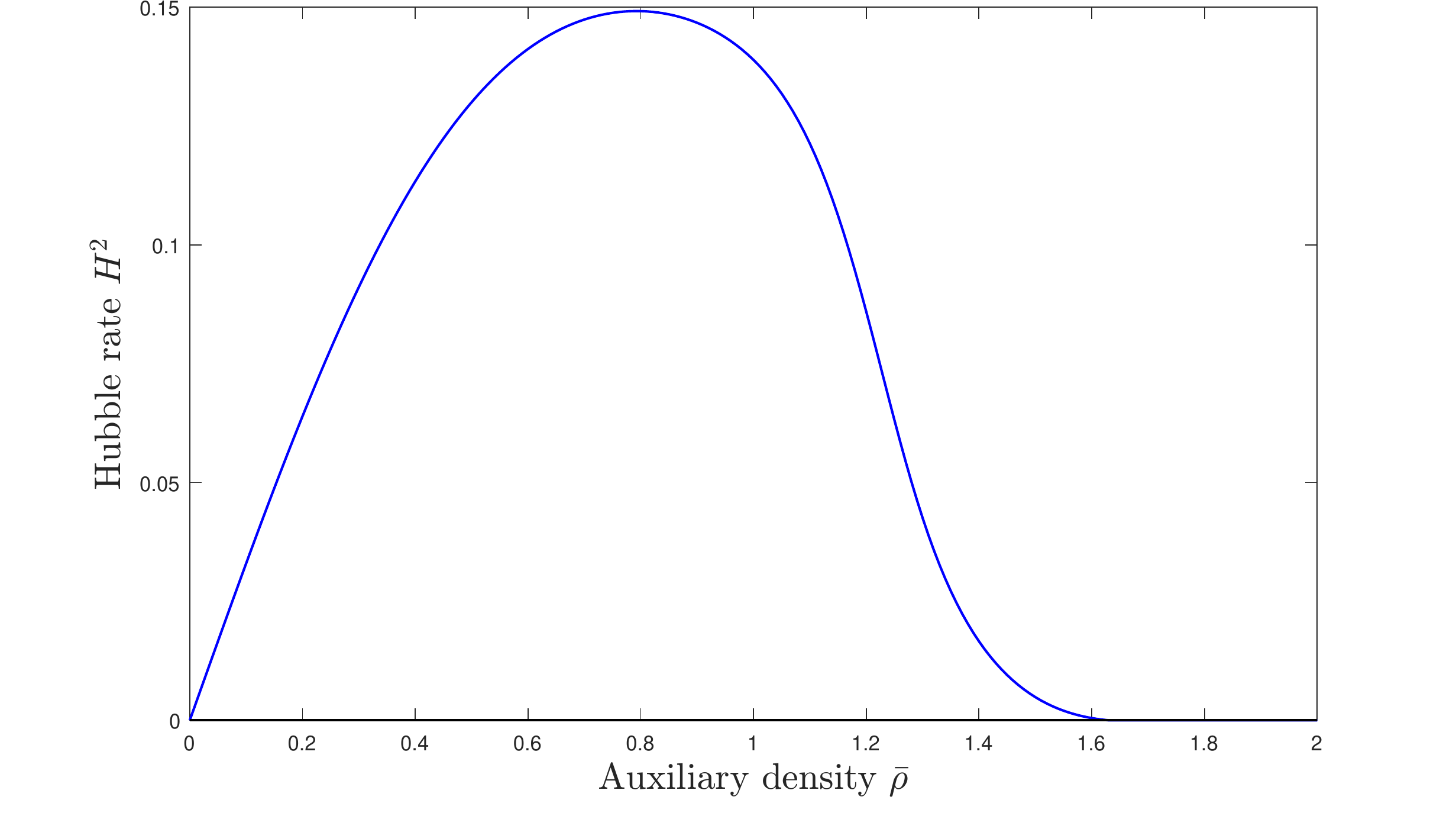}\includegraphics[width=0.5\textwidth]{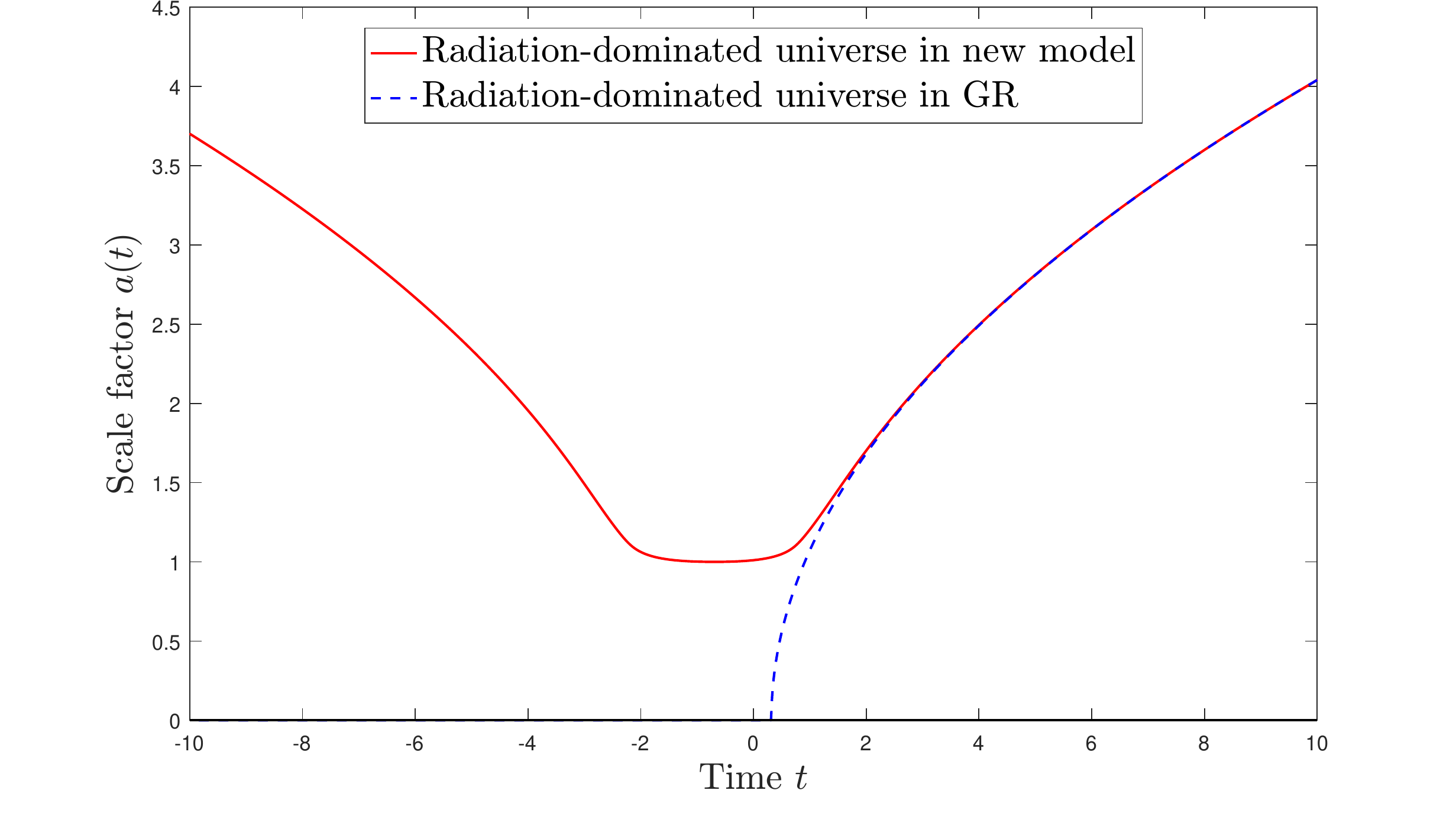}\\
  \caption{Left: Hubble rate $H^{2}$ against auxiliary density $\bar{\rho}$ in case (2)\\
  Right: Scale factor (normalized by minimum scale factor) against time in case (2). The red solid line represents the evolution in our model
  and the blue dotted line represents the evolution of GR.    }
  \label{figure2}
\end{figure}

\begin{figure}[h]
  \centering
  \includegraphics[width=0.5\textwidth]{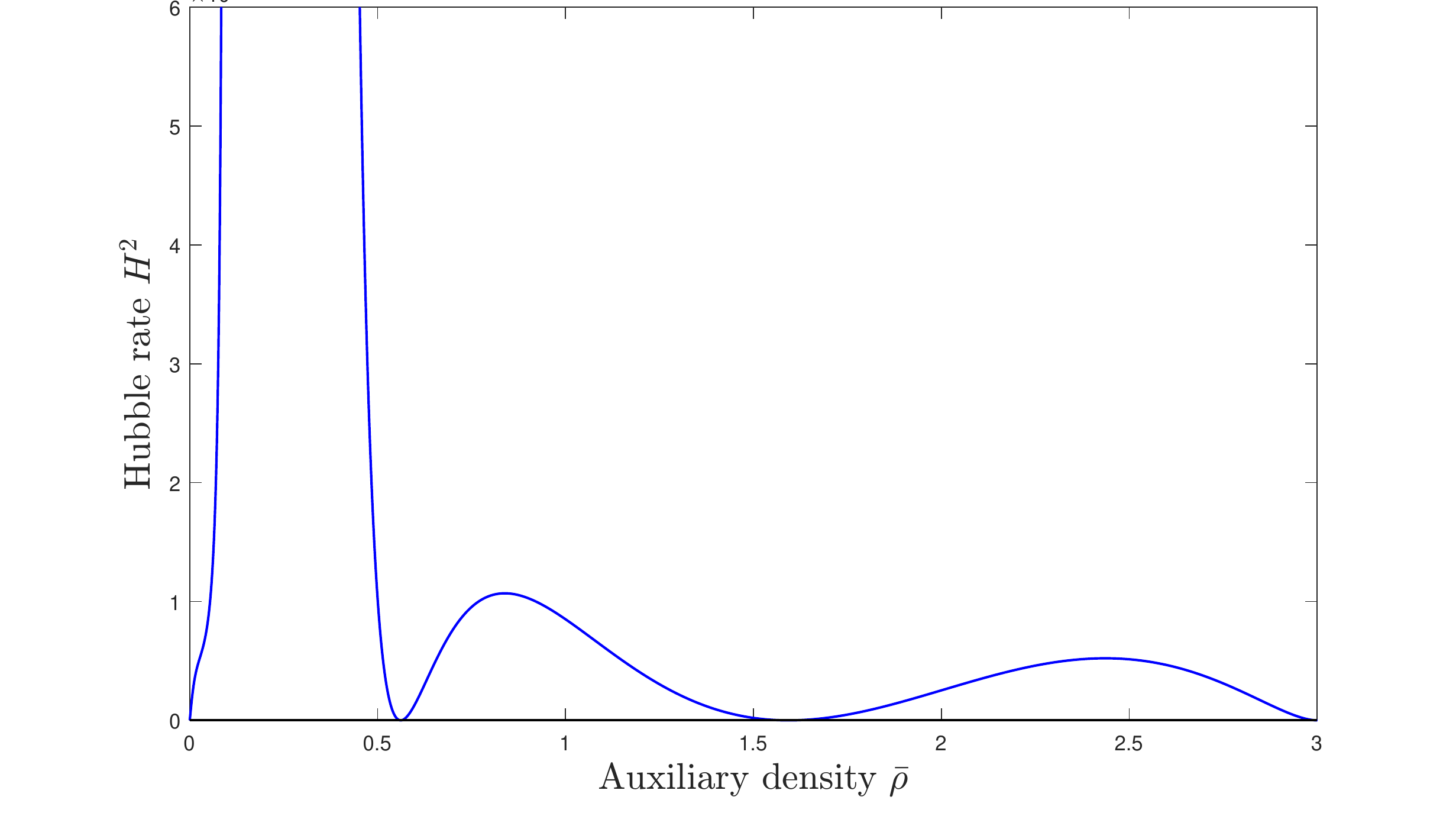}\includegraphics[width=0.5\textwidth]{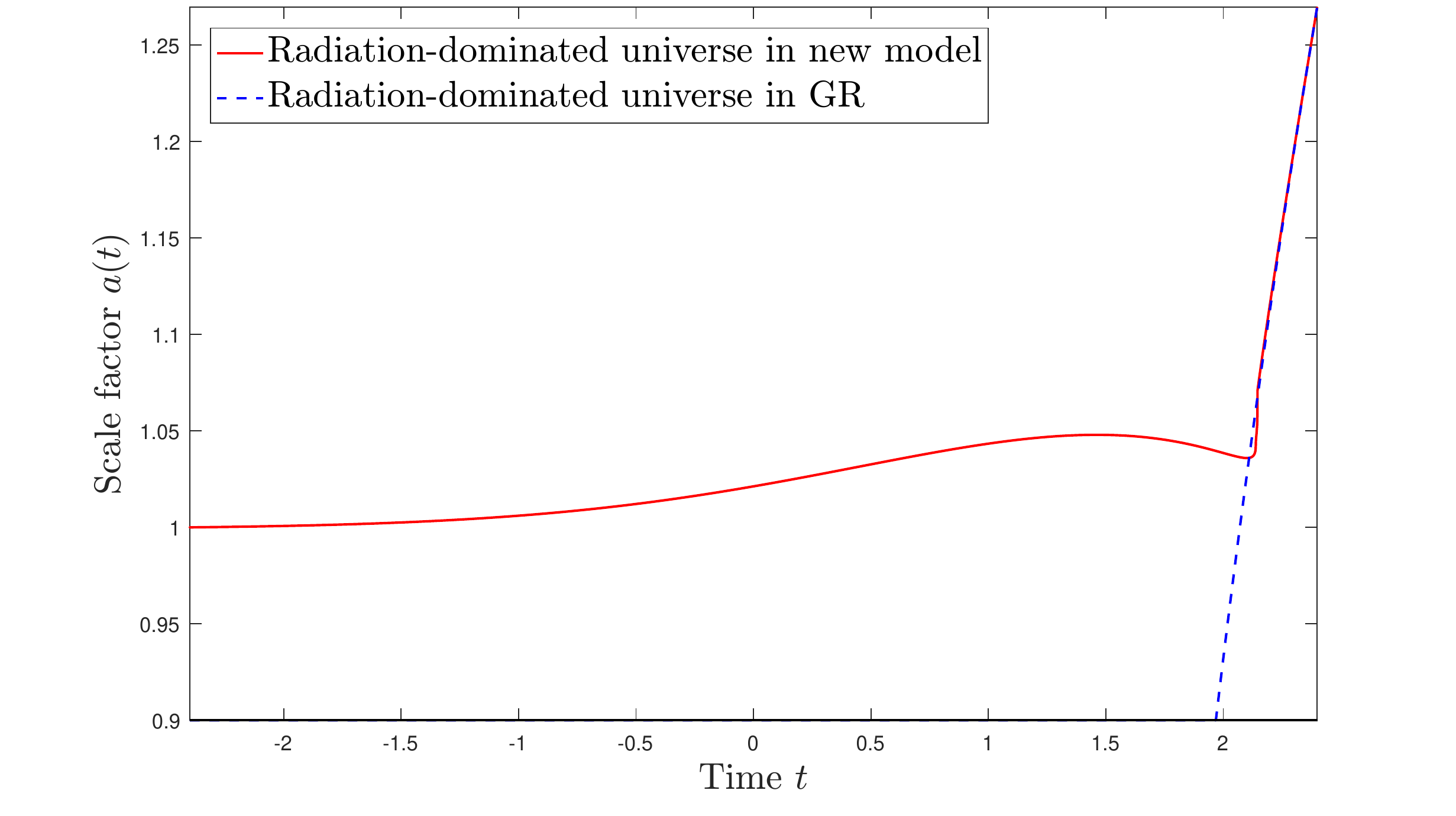}\\
  \caption{Left: Hubble rate $H^{2}$ against auxiliary density $\bar{\rho}$ in case (3) \\
  Right: Scale factor (normalized by minimum scale factor) against time in case (3). The red solid line represents the evolution in our model
  and the blue dotted line represents the evolution of GR.
    }
  \label{figure3}
\end{figure}

\begin{figure}[h]
  \centering
  \includegraphics[width=0.5\textwidth]{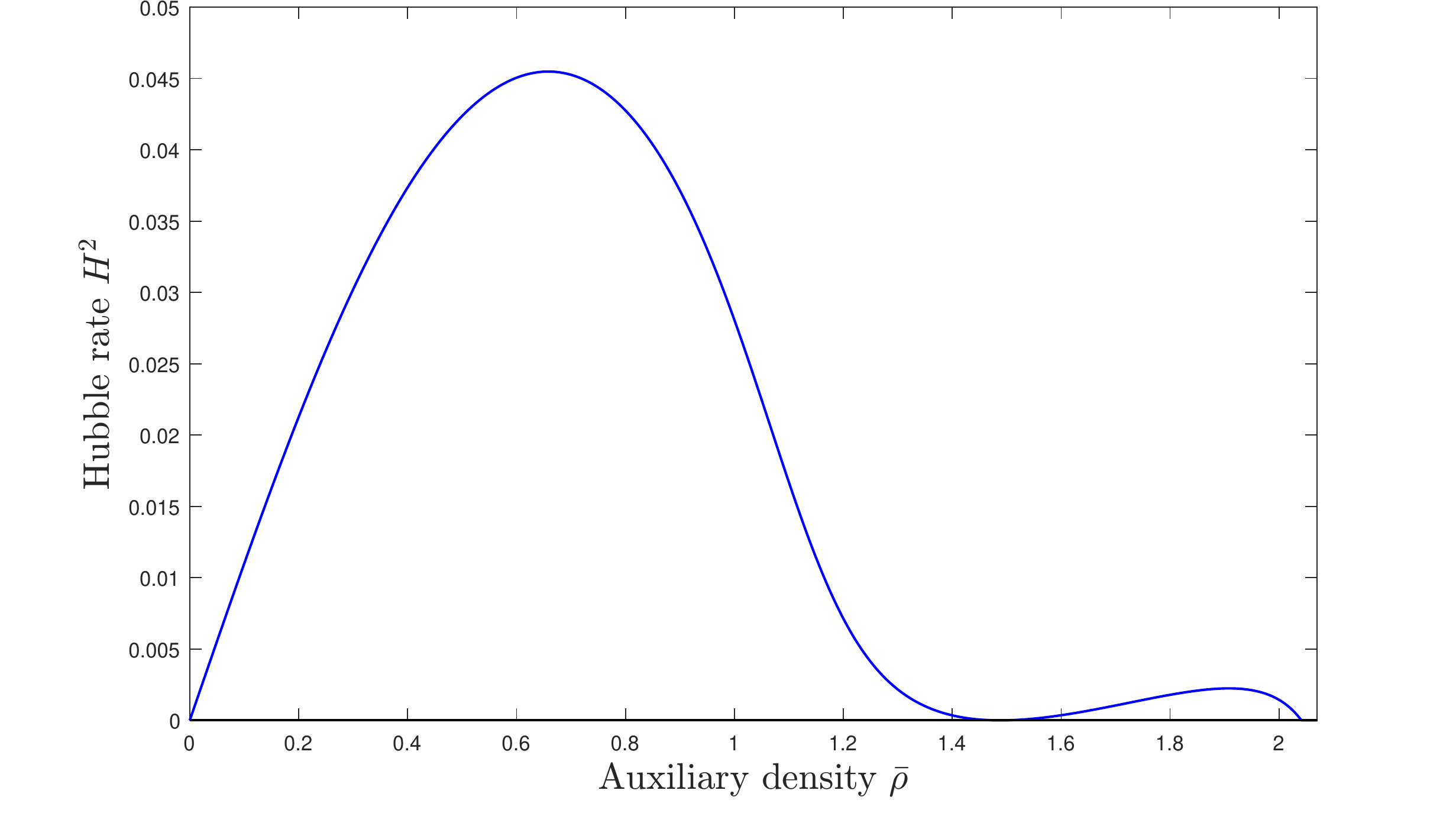}\includegraphics[width=0.5\textwidth]{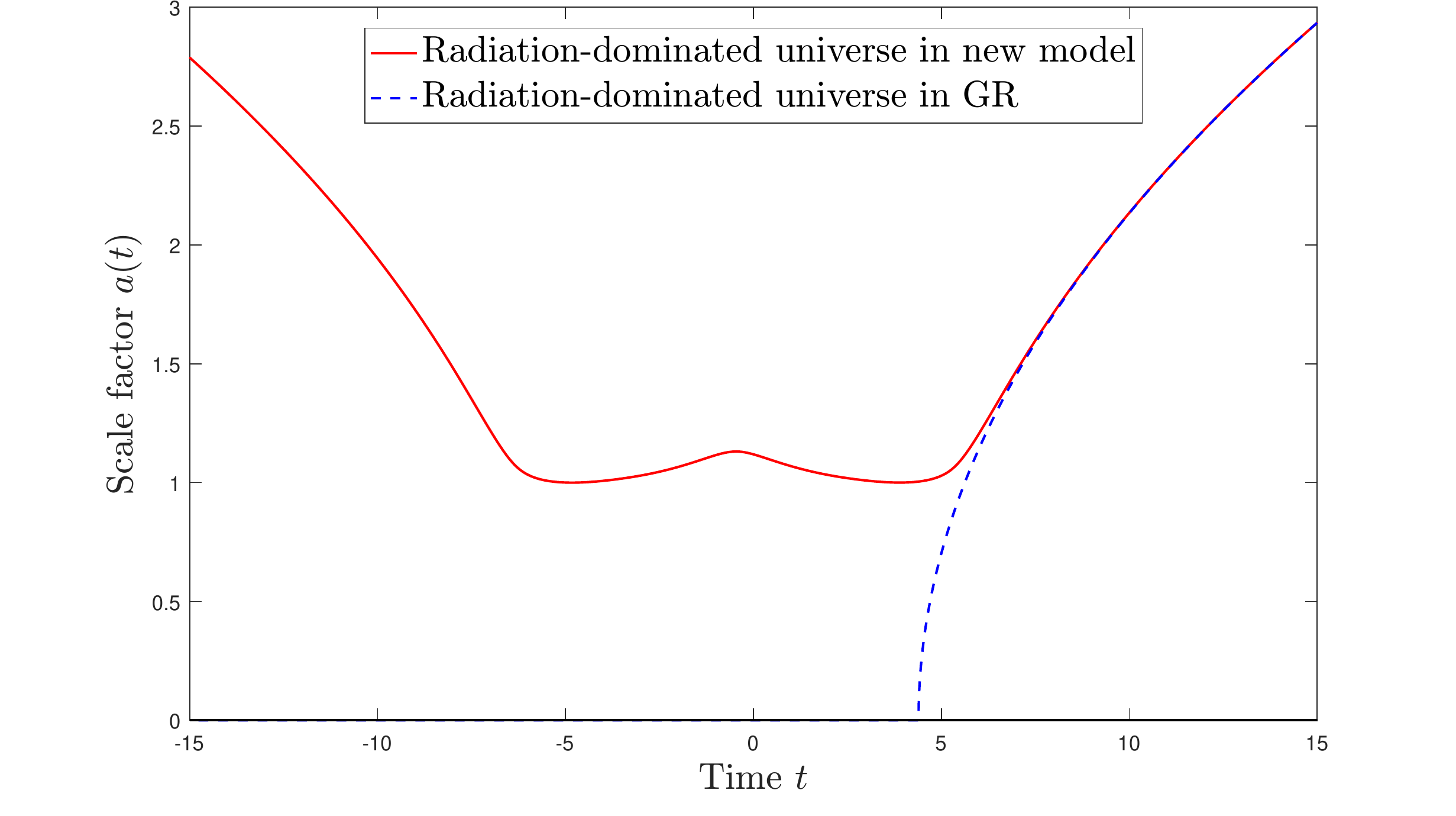}\\
  \caption{Left: Hubble rate $H^{2}$ against auxiliary density $\bar{\rho}$ in case (4)\\
  Right: Scale factor (normalized by minimum scale factor) against time in case (4). The red solid line represents the evolution in our model
  and the blue dotted line represents the evolution of GR. }
  \label{figure4}
\end{figure}

Let's briefly analyze what kinds of functions are appropriate.
For convenience, we take $8\pi G|_{\rho=0}=1$, this requires  $f_1(\kappa_0)=\kappa_0$,
where $\kappa=\kappa_{0}$ is the solution of (\ref{yueshu1}) when $\rho=0$.
In order to ensure that (\ref{Glimitation}) is satisfied at low energy density,
we take $(d\ln G/d\rho)|_{\rho=0}=0$, it is in turn requires $ f'_{1}(\kappa_0)=1$. This condition guarantees that at late time the rate of variation, $(d\ln G/dt)|_{t_0}$, is proportional to the second order derivative
$(d^2\ln G/d\rho^2)|_{\rho=0}$ and current Hubble constant $H_0$, both have extremely small values.
From constraint equation (\ref{constraint}), we can know that this condition also means that $(d\Lambda/d\rho)|_{\rho=0}=0$ or $ f'_{2}(\kappa_0)=\Lambda_0$,
where $\Lambda_0=\Lambda|_{\rho=0}$.

We first consider some examples where the functions $f_1$ and $f_2$ satisfy above conditions:

(1) $f_{1}=2e^{\kappa-3}+e^{3-\kappa}$ and $f_{2}=1/\cosh(\kappa-3)$. The behavior of the early universe is shown in FIG.\ref{figure1}.
We can see that, in this case, the scale factor remains as an almost constant at the early time and approaches to the standard big-bang at late time. This is similar to the case of $\kappa>0$ in the original EiBI model \cite{EIBI}.

(2) $f_1=\kappa^2-\kappa+1$ and $f_2=\kappa^2-2\kappa+2$.
The behavior of the early universe is shown in FIG.\ref{figure2}.
In this case there is a bounce in the early universe before entering into the standard big-bang cosmology.
 This behavior is similar to the behavior with $\kappa<0$ in the Ref. \cite{EIBI}.

(3) $f_1=\sinh(\kappa-10^{-2})+10^{-2}\cosh(\kappa-10^{-2})$ and $f_2=1/(1+(x-10^{-2})^2)$
The behavior of the early universe is shown in FIG.\ref{figure3}.
This case is similar to case (1) except that the scale factor oscillates slightly during the epoch before entering into the standard big-bang. The oscillating paradigm also appears in the relation between $H^2$ and $\bar{\rho}$.

(4) $f_1=2e^{\kappa-3}+e^{3-\kappa}$ and $f_{2}=\sqrt{\cosh(\kappa-3)}$.
The behavior of the early universe is shown in FIG.\ref{figure4}.
This case is similar to case (2) except that the scale factor oscillates slightly during the epoch before bouncing to the standard big-bang. The oscillating paradigm also appears in the relation between $H^2$ and $\bar{\rho}$.

In all above cases, the energy density of the universe has an upper bound, so that the scale factor has a lower bound and the singularity can be avoided  in the early universe.
And here we just give four examples, we can expect more different behaviors when choosing more different functions.

In the original EiBI model, there are only two different behaviors in a flat and radiation dominated universe \cite{EIBI}.
However, the early universe will have more different behaviors in the original EiBI model
if it is affected by other more strange components or by the spatial curvature \cite{EiBIcosmology}.
So, we can expect that in our model, the early universe will have more abundant behavior when considering the influence of different dominant components or spatial curvature.

\section{Conclusions}

In this paper, by replacing $\kappa$ and $\lambda$ in original EiBI action with the functions of $\kappa$,
we realize a varying cosmological constant in EiBI.
At the limit of small curvature, the theory returns to GR.
But Newton's constant and cosmological constant are variables,
and the changes are related to each other and are affected by the distribution of matter.
In order to be consistent with the experiment, we require that they change very little,
so there are some restrictions on the pending functions.
At the high curvature scales, the theory will seriously deviate from GR.
So the behavior of the early universe will be different from  standard cosmology.

Next, we study the cosmological behavior in a homogeneous isotropic background when the universe is dominated by radiation.
We find that the early universe will have different behaviors when we take different forms of pending functions.
Some of them behave similarly to the original EiBI model while others have different behaviors.
Same as the original EiBI model, all pending functions we have listed above can make the energy density has an upper bound,
so that the scale factor has a lower bound and the singularity can be avoided in the early universe.
And the late behavior is consistent with standard cosmology.

\section{Acknowledgement}

This work is supported in part by NSFC under Grant No. 11653002 and No. 11422543.

{}

\clearpage
\end{CJK*}

\begin{thebibliography}{}

 \bibitem{VaryingLambda}
  S.~Alexander, M.~Cort$\hat{\text{e}}$s, A.~R.~Liddle, J.~Magueijo, R.~Sims and L.~Smolin,
  Phys.\ Rev.\ D {\bf100} (2019) 083506 
  doi:10.1103/PhysRevD.100.083506
  [arXiv:1905.10380 [gr-qc]].

\bibitem{Alexander:2019wne}
  S.~Alexander, M.~Cort$\hat{\text{e}}$s, A.~R.~Liddle, J.~Magueijo, R.~Sims and L.~Smolin,
  Phys. Rev. D \textbf{100} (2019) 083507 
  doi:10.1103/PhysRevD.100.083507
  [arXiv:1905.10382 [gr-qc]].

\bibitem{eddington}
A.~S.~Eddington, The Mathematical Theory of Relativity, Cambridge University Press (1924); E.~Schrodinger, Spacetime Stucture, Cambridge University Press (1950).


\bibitem{EIBI}
 M.~Banados and P.~G.~Ferreira,
  Phys.\ Rev.\ Lett.\  {\bf 105} (2010) 011101
   Erratum: [Phys.\ Rev.\ Lett.\  {\bf 113} (2014) no.11,  119901]
  doi:10.1103/PhysRevLett.105.011101, 10.1103/PhysRevLett.113.119901
  [arXiv:1006.1769 [astro-ph.CO]].

\bibitem{Uzan:2010pm}
  J.~P.~Uzan,
  Living Rev.\ Rel.\  {\bf 14} (2011) 2
  doi:10.12942/lrr-2011-2
  [arXiv:1009.5514 [astro-ph.CO]].

\bibitem{Williams:2004qba}
  J.~G.~Williams, S.~G.~Turyshev and D.~H.~Boggs,
  Phys.\ Rev.\ Lett.\  {\bf 93} (2004) 261101
  doi:10.1103/PhysRevLett.93.261101
  [gr-qc/0411113].

\bibitem{Kaspi:1994hp}
  V.~M.~Kaspi, J.~H.~Taylor and M.~F.~Ryba,
  Astrophys.\ J.\  {\bf 428} (1994) 713.
  doi:10.1086/174280

\bibitem{Zhu:2018etc}
  W.~W.~Zhu {\it et al.},
  Mon.\ Not.\ Roy.\ Astron.\ Soc.\  {\bf 482} (2019) no.3,  3249
  doi:10.1093/mnras/sty2905
  [arXiv:1802.09206 [astro-ph.HE]].

\bibitem{Copi:2003xd}
  C.~J.~Copi, A.~N.~Davis and L.~M.~Krauss,
  Phys.\ Rev.\ Lett.\  {\bf 92} (2004) 171301
  doi:10.1103/PhysRevLett.92.171301
  [astro-ph/0311334].

\bibitem{Cyburt:2004yc}
  R.~H.~Cyburt, B.~D.~Fields, K.~A.~Olive and E.~Skillman,
  Astropart.\ Phys.\  {\bf 23} (2005) 313
  doi:10.1016/j.astropartphys.2005.01.005
  [astro-ph/0408033].

\bibitem{BI1988}
S.~Deser,  G.~W.~Gibbons,
Class.\ Quant.\ Grav.\ {\bf 15} (1998) L35
doi:10.1088/0264-9381/15/5/001
[arXiv:hep-th/9803049].

\bibitem{BI2004}
 D.~N.~Vollick,
 Phys.\ Rev.\ D {\bf 69} (2004) 064030
 doi:10.1103/PhysRevD.69.064030
 [arXiv:gr-qc/0309101].

\bibitem{BI2005}
D.~N.~Vollick,
Phys.\ Rev.\ D {\bf 72} (2005) 084026
doi:10.1103/PhysRevD.72.084026
[arXiv:gr-qc/0506091].

\bibitem{NewInsight}
 T.~Delsate, J.~Steinhoff,
Phys.\ Rev.\ Lett.\  {\bf 109} (2012) 021101
doi:10.1103/PhysRevLett.109.021101
[arXiv:1201.4989]

\bibitem{EiBIcosmology}
 J.H.C. Scargill, M. Ba$\tilde{\text{n}}$ados, P.G. Ferreira,
Phys.\ Rev.\ D {\bf 86} (2012) 103533
doi: 10.1103/PhysRevD.86.103533
[arXiv:1210.1521]



\end{thebibliography}
\end{document}